\documentclass[conference,a4paper]{IEEEtran}
\IEEEoverridecommandlockouts 
\usepackage{amsmath,graphicx}
\usepackage{cite}
\usepackage{balance}
\usepackage{multirow}
\usepackage{makecell}
\usepackage{pgfplots}
\usepackage{amssymb,amsfonts}
\usepackage{algorithmic}
\usepackage{xcolor}
\usepackage{boldline, soul}
\usepackage{booktabs}
\usepackage{indentfirst}
\let\labelindent\relax

\usepackage{enumitem}

\usepackage{pdfpages}
\usepackage{textcomp}

\usepackage{caption}

\def\BibTeX{{\rm B\kern-.05em{\sc i\kern-.025em b}\kern-.08em
    T\kern-.1667em\lower.7ex\hbox{E}\kern-.125emX}}
    
    \setcellgapes{3pt}
\def\BibTeX{{\rm B\kern-.05em{\sc i\kern-.025em b}\kern-.08em
    T\kern-.1667em\lower.7ex\hbox{E}\kern-.125emX}}

\pgfplotsset{compat=1.13}

\begin{document}

\title{Prediction-Aware Quality Enhancement of VVC Using CNN}

\author{

\IEEEauthorblockN{Fatemeh Nasiri $^{\star\dagger+}$, Wassim Hamidouche$^{\star\dagger}$, Luce Morin$^{\dagger\star}$, Nicolas Dhollande$^{+}$, Gildas Cocherel$^{+}$}

\IEEEauthorblockA{$^{\star}$ IRT b$<>$com, 35510 Cesson-S\'{e}vign\'{e}, France,\\
		 $^{\dagger}$ Univ Rennes, INSA Rennes, CNRS, IETR - UMR 6164, 35000 Rennes, France \\
		 $^+$ AVIWEST, 35760, Saint-Gr\'{e}goire, France
		 } 
}

\maketitle

\begin{abstract}
The upcoming video coding standard, Versatile Video Coding (VVC), has shown great improvement compared to its predecessor, High Efficiency Video Coding (HEVC), in terms of bitrate saving. Despite its substantial performance, compressed videos might still suffer from quality degradation at low bitrates due to coding artifacts such as blockiness, blurriness and ringing. In this work, we exploit Convolutional Neural Networks (CNN) to enhance quality of VVC coded frames after decoding in order to reduce low bitrate artifacts. The main contribution of this work is the use of coding information from the compressed bitstream. More precisely, the prediction information of intra frames is used for training the network in addition to the reconstruction information. The proposed method is applied on both luminance and chrominance components of intra coded frames of VVC. Experiments on VVC Test Model (VTM) show that, both in low and high bitrates, the use of coding information can improve the BD-rate performance by about 1\% and 6\% for luma and chroma components, respectively.
\end{abstract}


\section{Introduction}
\noindent
Video streaming applications have gained more popularity in the past few years. Therefore, the task of delivering a high quality video has become essential. From the compression point of view, the upcoming video coding standards, in particular VVC, can achieve up to $50\%$ bitrate saving compared to its predecessor HEVC~\cite{vvc}. Alongside the video coding progress, receiver devices have also become more powerful in processing received videos and enhancing their quality. As a result, video post-processing is nowadays an interesting option for display manufacturers in order to further improve the viewing experience of their users.

The promising performance of machine learning methods has recently encouraged researchers to exploit them in the video compression domain. Particularly, deep Convolutional Neural Networks (CNN) have attracted more attention owing to their significant performance \cite{rev1,rev2}. Despite the interesting performance of CNN-based methods, they usually impose a high computational complexity which makes them unsuitable for real-time encoding applications. However, the post-processing approaches which improve the reconstructed video after the decoding step can be more flexible, since they are not involved in the encoding and decoding process. In other words, Such post-processing approaches can serve as an optional step to be used based on the hardware capacity of the decoder device. 

CNN-based quality enhancement (QE) for VVC has been sparsely studied in the literature. The existing works target both intra and inter frames of coded videos. In \cite{JVET-N0254, JVET-O0079,JVET-M1001,JVET-N0513, JVET-N0710, lam2020efficient, meng2019enhancing, ma2020mfrnet}, CNN-based methods have been proposed to the VVC standardization either as in-loop filter or post-processing step. Considering the fact that the distortion in compressed video is influenced by the encoding process and its decision making engine, an attention based network is proposed in \cite{Attention-based}, where partitioning information of VVC is exploited to further increase the performance of the QE filter. Finally, in \cite{bullPP}, the impact of network architecture complexity on the performance of the QE filter has been studied. 

\begin{figure}[t]
    \centering
    \tikzset{every picture/.style={line width=0.75pt}} 

\begin{tikzpicture}[x=0.75pt,y=0.75pt,yscale=-1,xscale=1]

\draw    (216.33,125.6) -- (236.33,125.6) ;
\draw    (216.33,139.6) -- (236.33,139.6) ;

\draw [shift={(239.33,125.6)}, rotate = 180] [fill={rgb, 255:red, 0; green, 0; blue, 0 }  ][line width=0.08]  [draw opacity=0] (8.93,-4.29) -- (0,0) -- (8.93,4.29) -- cycle    ;

\draw [shift={(239.33,139.6)}, rotate = 180] [fill={rgb, 255:red, 0; green, 0; blue, 0 }  ][line width=0.08]  [draw opacity=0] (8.93,-4.29) -- (0,0) -- (8.93,4.29) -- cycle    ;

\draw (216.33,115.6) node [anchor=north west][inner sep=0.75pt]  [font=\tiny] [align=left] {Rec.};
\draw (216.33,144.6) node [anchor=north west][inner sep=0.75pt]  [font=\tiny] [align=left] {Pred.};

\draw    (46.33,132.6) -- (73.33,132.6) ;
\draw [shift={(76.33,132.6)}, rotate = 180] [fill={rgb, 255:red, 0; green, 0; blue, 0 }  ][line width=0.08]  [draw opacity=0] (8.93,-4.29) -- (0,0) -- (8.93,4.29) -- cycle    ;
\draw  [fill={rgb, 255:red, 255; green, 255; blue, 255 }  ,fill opacity=1 ] (131,114) -- (173.33,114) -- (173.33,154) -- (131,154) -- cycle ;
\draw  [fill={rgb, 255:red, 255; green, 255; blue, 255 }  ,fill opacity=1 ] (20,117) -- (46.83,117) -- (46.83,133.15) -- (20,133.15) -- cycle ;
\draw  [fill={rgb, 255:red, 255; green, 255; blue, 255 }  ,fill opacity=1 ] (23.83,121.04) -- (50.67,121.04) -- (50.67,137.19) -- (23.83,137.19) -- cycle ;
\draw  [fill={rgb, 255:red, 255; green, 255; blue, 255 }  ,fill opacity=1 ] (27.67,125.08) -- (54.5,125.08) -- (54.5,141.23) -- (27.67,141.23) -- cycle ;
\draw  [fill={rgb, 255:red, 255; green, 255; blue, 255 }  ,fill opacity=1 ] (31.5,129.11) -- (58.33,129.11) -- (58.33,145.27) -- (31.5,145.27) -- cycle ;

\draw  [fill={rgb, 255:red, 255; green, 255; blue, 255 }  ,fill opacity=1 ] (76,122.27) -- (98.33,122.27) -- (98.33,143.27) -- (76,143.27) -- cycle ;
\draw  [fill={rgb, 255:red, 128; green, 128; blue, 128 }  ,fill opacity=1 ] (143.17,139.05) .. controls (143.17,137.59) and (144.29,136.4) .. (145.67,136.4) .. controls (147.06,136.4) and (148.18,137.59) .. (148.18,139.05) .. controls (148.18,140.51) and (147.06,141.7) .. (145.67,141.7) .. controls (144.29,141.7) and (143.17,140.51) .. (143.17,139.05) -- cycle ;
\draw    (152.97,133.53) -- (147.63,137.09) ;
\draw  [fill={rgb, 255:red, 128; green, 128; blue, 128 }  ,fill opacity=1 ] (152.32,132.15) .. controls (152.32,130.68) and (153.44,129.5) .. (154.83,129.5) .. controls (156.21,129.5) and (157.33,130.68) .. (157.33,132.15) .. controls (157.33,133.61) and (156.21,134.79) .. (154.83,134.79) .. controls (153.44,134.79) and (152.32,133.61) .. (152.32,132.15) -- cycle ;
\draw  [fill={rgb, 255:red, 128; green, 128; blue, 128 }  ,fill opacity=1 ] (152.32,145.95) .. controls (152.32,144.49) and (153.44,143.31) .. (154.83,143.31) .. controls (156.21,143.31) and (157.33,144.49) .. (157.33,145.95) .. controls (157.33,147.42) and (156.21,148.6) .. (154.83,148.6) .. controls (153.44,148.6) and (152.32,147.42) .. (152.32,145.95) -- cycle ;
\draw    (147.53,140.43) -- (152.87,143.88) ;

\draw  [fill={rgb, 255:red, 184; green, 233; blue, 134 }  ,fill opacity=1 ] (240,119.27) -- (274.33,119.27) -- (274.33,149.27) -- (240,149.27) -- cycle ;
\draw    (98.33,132.6) -- (127.33,132.6) ;
\draw [shift={(130.33,132.6)}, rotate = 180] [fill={rgb, 255:red, 0; green, 0; blue, 0 }  ][line width=0.08]  [draw opacity=0] (8.93,-4.29) -- (0,0) -- (8.93,4.29) -- cycle    ;
\draw    (173.33,132.6) -- (192.33,132.6) ;
\draw [shift={(195.33,132.6)}, rotate = 180] [fill={rgb, 255:red, 0; green, 0; blue, 0 }  ][line width=0.08]  [draw opacity=0] (8.93,-4.29) -- (0,0) -- (8.93,4.29) -- cycle    ;

\draw    (205,122.27) -- (205,107) ;
\draw    (205,107) -- (257,107) ;
\draw    (257,107) -- (257,119.27) ;
\draw [shift={(257,119.27)}, rotate = 180] [fill={rgb, 255:red, 0; green, 0; blue, 0 }  ][line width=0.08]  [draw opacity=0] (0,0) -- (-4.29,8.93) -- (4.29,8.93) -- cycle    ;
\draw (226,98) node [anchor=north west][inner sep=0.75pt]  [font=\tiny] [align=left] {QP};

\draw    (275.33,132.6) -- (295.33,132.6) ;
\draw [shift={(298.33,132.6)}, rotate = 180] [fill={rgb, 255:red, 0; green, 0; blue, 0 }  ][line width=0.08]  [draw opacity=0] (8.93,-4.29) -- (0,0) -- (8.93,4.29) -- cycle    ;
\draw  [fill={rgb, 255:red, 255; green, 255; blue, 255 }  ,fill opacity=1 ] (299,122) -- (325.83,122) -- (325.83,138.15) -- (299,138.15) -- cycle ;
\draw  [fill={rgb, 255:red, 255; green, 255; blue, 255 }  ,fill opacity=1 ] (302.83,126.04) -- (329.67,126.04) -- (329.67,142.19) -- (302.83,142.19) -- cycle ;
\draw  [fill={rgb, 255:red, 255; green, 255; blue, 255 }  ,fill opacity=1 ] (306.67,130.08) -- (333.5,130.08) -- (333.5,146.23) -- (306.67,146.23) -- cycle ;
\draw  [fill={rgb, 255:red, 255; green, 255; blue, 255 }  ,fill opacity=1 ] (310.5,134.11) -- (337.33,134.11) -- (337.33,150.27) -- (310.5,150.27) -- cycle ;

\draw  [fill={rgb, 255:red, 255; green, 255; blue, 255 }  ,fill opacity=1 ] (194,122.27) -- (216.33,122.27) -- (216.33,143.27) -- (194,143.27) -- cycle ;
\draw  [color={rgb, 255:red, 255; green, 255; blue, 255 }  ,draw opacity=1 ][fill={rgb, 255:red, 255; green, 255; blue, 255 }  ,fill opacity=1 ] (100,159.27) -- (170,159.27) -- (170,170) -- (100,170) -- cycle ;

\draw (77,127) node [anchor=north west][inner sep=0.75pt]  [font=\scriptsize] [align=left] {Enc.};
\draw (246,127) node [anchor=north west][inner sep=0.75pt]   [align=left] {QE};
\draw (133,117) node [anchor=north west][inner sep=0.75pt]  [font=\scriptsize] [align=left] {Channel};
\draw (100.33,121.27) node [anchor=north west][inner sep=0.75pt]  [font=\tiny] [align=left] {Stream};
\draw (195,127) node [anchor=north west][inner sep=0.75pt]  [font=\scriptsize] [align=left] {Dec.};
\draw (296,106) node [anchor=north west][inner sep=0.75pt]  [font=\scriptsize] [align=left] {Display};
\draw (17,100) node [anchor=north west][inner sep=0.75pt]  [font=\scriptsize] [align=left] {Capture};

\end{tikzpicture}
    \caption{Compressed video quality enhancement framework}
    \label{framework}
\end{figure}
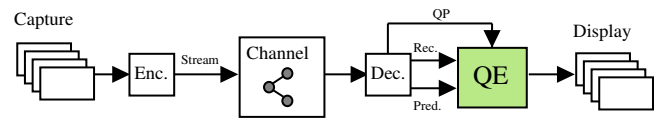

In this paper, a CNN-based QE method is proposed, which follows the objective of the previously presented works with the use of coding information~\cite{JVET-N0169, JVET-N0254, Attention-based}. The main contribution of this work is that we use the spatial predictor of each frame as the input to the CNN. This is motivated by the fact that coding information, such as intra prediction signal, usually represent a useful information about the type of the distortion \cite{ipta}. Fig~\ref{framework} presents the overall workflow of the proposed method. The input of the QE neural network is the decoded frame, the intra prediction information and the Quantization Parameter (QP). The CNN architecture of this paper is inspired by the network proposed in \cite{EDSR}, which has shown great performance for the super resolution problem. Moreover, the three color components of each frame are processed separately. 

The rest of this paper is organized as follows. In Section~\ref{Sec:PredInf}, the proposed QE method using intra prediction as coding information is presented. Experimental results as well as discussions and comparisons with state of the art solutions are provided in Section~\ref{sec:res} and finally, Section \ref{sec:con} concludes the paper.

\IEEEpubidadjcol

\section{Predictor-aware quality enhancement}
\label{Sec:PredInf}
\noindent
In this section, first we will explain the intuition and motivation for using intra prediction in the proposed CNN-based QE method. Then, network architecture and training configuration will be presented. 

\begin{figure}
    \centering
     \begin{center}
     \begin{tabular}{ cccl  }
    
    \multicolumn{4}{c}{Selected block ($\mathcal{O}^k_i$)}
    \\ 
    \multicolumn{4}{c}{\includegraphics[scale=0.72]{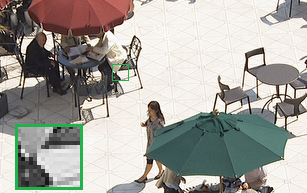}}
    \\
    
    & & & 
    
    \\
    
    \scriptsize{Prediction ($\mathcal{P}^k_i$)} & \scriptsize{Reconstruction ($\mathcal{C}^k_i$)} & \scriptsize{Loss ($\mathcal{O}^k_i$ - $\mathcal{C}^k_i$)} & 
    \\
    
    \multirow{4}{*}{\includegraphics[scale=0.35]{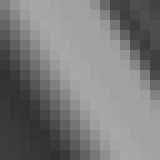} }
    & 
    \multirow{4}{*}{\includegraphics[scale=0.35]{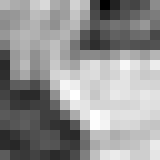} }
    & 
    \multirow{4}{*}{\includegraphics[scale=0.35]{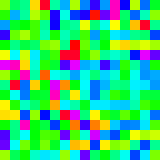}}
    & 
    \scriptsize{IPM 38}
    
    \\
    & & & \scriptsize{R$_{38}$: 182}
    \\ 
    & & & \scriptsize{D$_{38}$.: 22970}
    \\
    & & & \scriptsize{J$_{38}$: 77803}
    \\ 
    & & &
    \\
    \multirow{4}{*}{\includegraphics[scale=0.35]{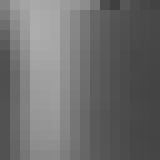} }
    & 
    \multirow{4}{*}{\includegraphics[scale=0.35]{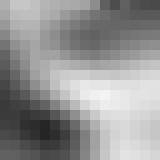} }
    & 
    \multirow{4}{*}{\includegraphics[scale=0.35]{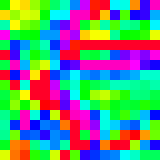}}
    &
    \scriptsize{IPM 50}
    
    \\
    & & & \scriptsize{R$_{50}$: 112}
    \\ 
    & & & \scriptsize{D$_{50}$.: 43518}
    \\
    & & & \scriptsize{J$_{50}$: 77390}
    \\ 
    & & &
      
    \end{tabular}
    \end{center}
    \caption{A 16 $\times$ 16 block, \textit{k}, and its two best IPMs (\textit{i} = 38, 50), with similar costs but different rate-distortion trade-offs resulting in distinct compression loss patterns (QP 40: $\lambda$=301)}
    \label{intraprediction}
\end{figure}

\subsection{Intra coding and compression artifacts}

\noindent
In intra coding, each block is predicted based on its neighboring pixels, given some predefined models. In VVC, these models include a set of 67 Intra Prediction Modes (IPM), representing 65 angular IPMs, plus DC and planar. Like other decisions in video coding, the selection of an IPM for a block consists in optimizing a function of the rate and the distortion, called the rate-distortion (R-D) cost. Particularly for intra coding, the the R-D cost of an IPM \textit{i}, denoted as $J_i$, is computed as    
\begin{equation}
    J_i = D_i+\lambda \times R_i \text{\hspace{15pt} } i=1,...,67,
    \label{RD}
\end{equation}
where $D_i$ and $R_i$ are the distortion and the rate, obtained when using \textit{i} as the IPM of the block, respectively. Moreover, $\lambda$ is the Lagrangian multiplier, computed based on the QP which determines the relative importance of the rate and the distortion during the decision making process. For instance, in low bitrates (high QP), the value of $\lambda$ is higher, which indicates that minimization of the rate is relatively more important than minimization of the distortion. 

\begin{figure*}
    \centering
    \includegraphics[scale=.25]{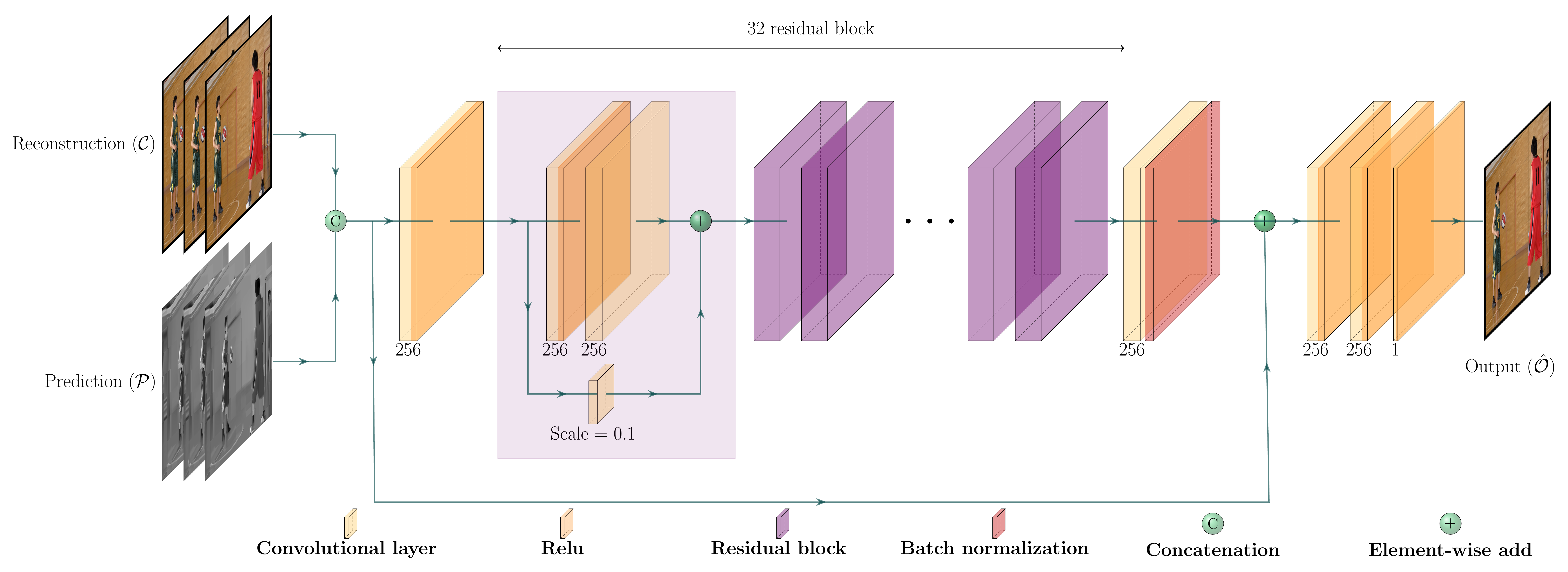}
    \caption{Network architecture of the proposed method using the prediction and the reconstruction signal as the input.}
    \label{arch}
\end{figure*}

Strict bitrate constraints might cause a situation where the best IPM minimizing the R-D cost of a block, is not necessarily the IPM that models the block texture most accurately. Fig.~\ref{intraprediction} shows an example of such a situation in the first frame of the BQSquare sequence. In this figure, a $16\times16$ block, \textit{k}, is selected and the Prediction ($\mathcal{P}_i^k$) and Reconstruction ($\mathcal{C}_i^k$) blocks corresponding to its two best IPMs in terms of R-D cost are shown. As can be seen, despite their similar R-D costs, these two IPMs result in very different reconstruction signals, with different types of compression loss patterns. This behavior is due to two different R-D trade offs of the selected modes.

On one hand, IPM 38 is able to model the block content more accurately (i.e. smaller distortion $D_{38}$) with the cost of a higher IPM and residual coding rate (i.e. $R_{38}$). On the other hand, IPM 50 provides a less accurate texture modeling (i.e. high distortion $D_{50}$) with a smaller rate residual and IPM coding rate (i.e. $R_{50}$). Consequently, these two IPMs result in very different types of artifacts for the given block, as can be seen by comparing the corresponding reconstruction blocks (i.e. $\mathcal{C}_{38}^k$ and $\mathcal{C}_{50}^k$).

The above example proves that the task of QE for a block, frame or an entire sequence could be significantly impacted by different choices of coding modes (e.g. IPM) determined by the encoder. This assumption is the main motivation in our work to use the intra prediction information for training of the quality enhancement networks.

\subsection{Proposed CNN-based quality enhancement method}
\noindent
The proposed QE algorithm is applied on intra frames after decoding. In order to accurately capture the compression loss, as explained in previous section, the prediction information is also extracted from the decoder and is used as the input to the QE network. For each reconstruction frame, this prediction information is composed of predictors associated to its blocks. The predictor of each block is the projection of its reference pixels corresponding to the angle of the used IPM. The reconstruction and prediction frames are concatenated and fed to the network as one input image. 

Inspired by the architecture of the Enhanced Deep Super Resolution (EDSR)~\cite{EDSR}, we have exploited residual training in our QE network. The architecture of the QE network is shown in Fig~\ref{arch}. The first convolutional layer receives the reconstruction and prediction frames as input. In the next step, after one convolutional layer, 32 identical residual blocks, each composed of two convolutional layers, and one Relu layer in between, are used. The convolutional layers in the residual blocks have the same size as the feature maps and kernel size of first convolutional layer. In order to normalize the feature maps, a convolutional layer with batch normalization is applied after the residual blocks. A skip connection between the input of the first and the last residual block is used. Two more convolutional layers after the residual blocks are used. Finally, the last convolutional layer has one feature map which constructs the output frame. 

Given $\mathcal{I=P\oplus C}$ as the concatenation of the prediction $\mathcal{P}$ and reconstruction $\mathcal{C}$ frames as input, producing the enhanced frame $\mathcal{\hat{O}}$, is formulated as

\begin{equation}
\mathcal{\hat{O}}= F_1^1(F_2^2(Bn(F^1_3(Res^{32}(F_2^1(\mathcal{I})))) + \mathcal{I})),
\end{equation}
where $F_1(.)$ and $F_2(.)$ are $3\times 3 \times 256$ convolutional layer, with and without the Relu activation layer, respectively. Moreover, $F_3(.)$ is a $3\times 3 \times 1$ convolutional layer with Relu activation layer. The superscript of each function indicates the number of times they are repeated sequentially in the network architecture. Finally, $Res$ and $Bn$ are the residual block and batch normalization layer, respectively.

The task of the training phase is to optimize the parameters $\theta_{QE}$ of the above QE function, $f_{QE}$, expressed as
\begin{equation}
\mathcal{\hat{O}}=f_{QE}(\mathcal{C}, \mathcal{P}; \theta_{QE}).
\end{equation}

The $L_2$ norm with respect to the original frame $\mathcal{O}$ is used as the cost function of the training phase
\begin{equation}
    L_2(\mathcal{O}-\mathcal{\hat{O}})=||\mathcal{O}-\mathcal{\hat{O}}||^2_2.
\end{equation}

In the proposed method, each color component of the decoded video  (i.e. one luminance and two chrominance) is enhanced separately. For this purpose, one network for each component in different QPs is trained with the above network architecture and using the corresponding prediction signal of that component.


\section{Experimental Results}
\label{sec:res}
\noindent
As the proposed post-processing module is designed to enhance the quality of intra coded frames, two image datasets of DIV2K and Fliker2K are used for training. All images in the datasets are encoded in All-Intra (AI) configuration of the VVC Test Model version 5.0 (VTM-5.0) \cite{vvc}, using 6 QPs, between 22 and 47. The prediction information is extracted during decoding process for all datasets in all QP ranges.

The network was implemented and trained in pyTorch (1.4.0). For training, 64 $\times$ 64 patches of reconstruction and prediction frames were extracted randomly from the training dataset, with batch size of 32. The training started with the learning rate of $10^{-4}$ which was then decayed by the scale of 0.1 for each 100 epochs until 500 epochs. At the end of the training, a total of 3$\times$6 trained models were obtained for 3 components in 6 QPs. 

In order to evaluate our method, the test sequences of JVET CTC (classes A1, A2, B, C, D, E) were encoded with the VTM-5.0 with each of the 6 QPs with the AI configuration. To study the effect of QE method in different bit-rates, two QP ranges were evaluated: 1) the CTC QP: (22, 27, 32, 37), and 2) high QP: (32, 37, 42, 47). The performance of different benchmark methods were measured using the Bjontegaard delta (BD) bit-rate saving metric based on the PSNR difference with respect to VTM-5.0 with no QE as an anchor.  

Three state of the art VVC CNN-based QE methods are used as benchmark. First two methods are JVET contributions \cite{JVET-N0254, JVET-N0169} proposing QE methods as post processing. Both of these methods deploy a slightly simpler network architecture with the QP map and use the reconstruction signal as the only input to the network. To assess the benefit of using IPM as input to the network, we also present the results for the proposed method with only reconstruction frame as input (denoted "proposed - without prediction")

\begin{table*}[!ht]

\caption{Performance comparison in percentage (\%) of the proposed method against the VVC in terms of BD-Rate.}
\label{tabbdr2}
\centering
\scalebox{0.95}{

\begin{tabular}{clccc|ccc|ccc|ccc||ccc|ccc}
\hline
\hline							
 \multirow{4}{*}{Class} & \multirow{4}{*}{Sequence} & \multicolumn{12}{c||}{CTC QP (22-37)}	& \multicolumn{6}{c}{High QP (32-47)}\\ \cline{3-20}
                       &                     & \multicolumn{3}{c|}{{JVET-N0254 [4]}}                                               & \multicolumn{3}{c|}{{JVET-N0169 [10]}}                                               & \multicolumn{6}{c||}{Proposed {(VTM 5.0)}}                                           & \multicolumn{6}{c}{Proposed (VTM 5.0)}                                                 \\

                       &                     & \multicolumn{3}{c|}{(VTM 4.0)}                                                                               & \multicolumn{3}{c|}{(VTM 4.0)}                                                                               & \multicolumn{3}{c|}{Without prediction}                                  & \multicolumn{3}{c||}{With Prediction}                                    & \multicolumn{3}{c|}{Without prediction}                                  & \multicolumn{3}{c}{With Prediction} \\

\cline{3-20} 
                       &                     & Y                                  & U                                  & V                                  & Y                                  & U                                  & V                                  & Y                      & U                      & V                      & Y                      & U                      & V                      & Y                      & U                      & V                      & Y                      & U                      & V                      \\
\hline
\multirow{4}{*}{A1}    & Tango               & -0.9                               & -2.7                               & -3.3                               & -3.7                               & -7.8                               & -8.1                               & -4.3                   & -4.4                   & -9.1                   & -5.4                   & -21.8                  & -21.1                  & -6.4                   & -10.5                  & -11.9                  & -7.9                   & -20.2                  & -16.9                  \\  
                       & FoodMarket          & -1.3                               & -1.7                               & -2.3                               & -3.8                               & -3.8                               & -4.1                               & -9.9                   & -3.8                   & -9.7                   & -11.3                  & -12.2                  & -13.3                  & -7.8                   & -8.9                   & -10.9                  & -8.8                   & -12.8                  & -11.0                  \\  
                       & CampFire            & -0.6                               & -0.6                               & -3.1                               & -2.5                               & -9.5                               & -8.3                               & -3.3                   & -3.2                   & -11.1                  & -4.0                   & -8.2                   & -20.2                  & -6.4                   & -8.3                   & -11.2                  & -7.6                   & -10.0                  & -18.8                  \\  
                       & \textbf{Average}    & \textbf{-1.0}                      & \textbf{-1.6}                      & \textbf{-2.9}                      & \textbf{-3.3}                      & \textbf{-7.0}                      & \textbf{-6.8}                      & \textbf{-5.8}          & \textbf{-3.8}          & \textbf{-10.0}         & \textbf{-6.9}          & \textbf{-14.1}         & \textbf{-18.2}         & \textbf{-6.9}          & \textbf{-9.3}          & \textbf{-11.3}         & \textbf{-8.1}          & \textbf{-14.4}         & \textbf{-15.6}         \\ 
\hline
\multirow{4}{*}{A2}    & CatRobot            & -2.2                               & -3.7                               & -3.8                               & -4.6                               & -7.3                               & -10.8                              & -5.7                   & -4.7                   & -11.3                  & -6.4                   & -14.0                  & -18.8                  & -6.7                   & -13.1                  & -14.4                  & -8.0                   & -18.0                  & -17.7                  \\  
                       & Daylight            & -1.1                               & -4.3                               & -1.2                               & -3.5                               & -4.6                               & -5.9                               & -2.7                   & -1.7                   & -5.3                   & -3.7                   & -11.9                  & -11.5                  & -7.4                   & -7.9                   & -6.8                   & -8.9                   & -11.9                  & -9.2                   \\  
                       & ParkRunning         & -1.1                               & -0.2                               & -0.4                               & -3.6                               & -3.7                               & -3.4                               & -3.8                   & -0.6                   & -1.6                   & -4.4                   & -4.5                   & -4.4                   & -4.6                   & -2.2                   & -2.3                   & -5.6                   & -5.6                   & -5.3                   \\  
                       & \textbf{Average}    & \textbf{-1.5}                      & \textbf{-2.7}                      & \textbf{-1.8}                      & \textbf{-3.9}                      & \textbf{-5.2}                      & \textbf{-6.7}                      & \textbf{-4.1}          & \textbf{-2.3}          & \textbf{-6.1}          & \textbf{-4.8}          & \textbf{-10.1}         & \textbf{-11.6}         & \textbf{-6.2}          & \textbf{-7.8}          & \textbf{-7.8}          & \textbf{-7.5}          & \textbf{-11.8}         & \textbf{-10.7}         \\ 
\hline
\multirow{6}{*}{B}     & MarketPlace         & -0.9                               & -2.6                               & -2.9                               & -3.6                               & -3.7                               & -3.6                               & -4.2                   & -3.3                   & -7.9                   & -5.2                   & -13.6                  & -14.1                  & -5.2                   & -10.1                  & -11.3                  & -6.5                   & -17.6                  & -15.9                  \\  
                       & RitualDance         & -2.2                               & -2.3                               & -4.5                               & -6.0                               & -3.7                               & -4.5                               & -8.9                   & -4.3                   & -11.4                  & -10.2                  & -14.2                  & -16.9                  & -7.4                   & -9.7                   & -13.8                  & -8.9                   & -17.0                  & -17.4                  \\  
                       & Cactus              & -0.7                               & -2.5                               & -0.8                               & -4.0                               & -4.2                               & -9.1                               & -4.2                   & -2.1                   & -10.9                  & -4.9                   & -10.2                  & -15.7                  & -6.9                   & -9.9                   & -15.2                  & -8.0                   & -15.3                  & -18.0                  \\  
                       & BasketballDrive     & -0.3                               & -2.3                               & -3.5                               & -4.2                               & -11.8                              & -14.7                              & -5.1                   & -8.2                   & -16.1                  & -6.2                   & -18.9                  & -21.0                  & -6.3                   & -12.8                  & -17.4                  & -7.7                   & -17.9                  & -19.4                  \\  
                       & BQTerrace           & -0.5                               & -1.2                               & -3.7                               & -2.3                               & -7.2                               & -6.7                               & -3.4                   & -4.3                   & -8.2                   & -3.9                   & -13.6                  & -12.5                  & -7.5                   & -9.9                   & -9.4                   & -8.7                   & -15.3                  & -13.7                  \\  
                       & \textbf{Average}    & \textbf{-0.9}                      & \textbf{-2.2}                      & \textbf{-3.1}                      & \textbf{-4.0}                      & \textbf{-6.1}                      & \textbf{-7.7}                      & \textbf{-5.2}          & \textbf{-4.5}          & \textbf{-10.9}         & \textbf{-6.1}          & \textbf{-14.1}         & \textbf{-16.0}         & \textbf{-6.7}          & \textbf{-10.5}         & \textbf{-13.4}         & \textbf{-8.0}          & \textbf{-16.6}         & \textbf{-16.9}         \\ 
\hline
\multirow{5}{*}{C}     & BasketballDrill     & -3.0                               & -3.2                               & -5.2                               & -8.1                               & -14.8                              & -20.5                              & -9.1                   & -11.8                  & -24.3                  & -10.3                  & -23.3                  & -28.0                  & -8.7                   & -19.2                  & -21.2                  & -10.2                  & -21.2                  & -23.0                  \\  
                       & BQMall              & -2.1                               & -2.5                               & -4.3                               & -6.3                               & -5.9                               & -7.1                               & -6.7                   & -4.5                   & -7.6                   & -7.4                   & -11.7                  & -11.8                  & -7.6                   & -10.4                  & -11.3                  & -8.7                   & -16.9                  & -15.9                  \\  
                       & PartyScene          & -1.8                               & -1.3                               & -1.5                               & -4.2                               & -4.5                               & -4.7                               & -4.4                   & -3.1                   & -6.0                   & -4.8                   & -8.5                   & -10.0                  & -6.6                   & -9.1                   & -9.7                   & -7.5                   & -14.8                  & -15.3                  \\  
                       & RaceHorses          & -0.7                               & -2.4                               & -2.4                               & -3.6                               & -5.9                               & -9.3                               & -3.3                   & -3.6                   & -9.2                   & -3.8                   & -8.1                   & -12.3                  & -4.8                   & -14.0                  & -18.4                  & -5.7                   & -20.4                  & -22.8                  \\  
                       & \textbf{Average}    & \textbf{-1.9}                      & \textbf{-2.3}                      & \textbf{-3.3}                      & \textbf{-5.6}                      & \textbf{-7.8}                      & \textbf{-10.4}                     & \textbf{-5.9}          & \textbf{-5.7}          & \textbf{-11.8}         & \textbf{-6.6}          & \textbf{-12.9}         & \textbf{-15.5}         & \textbf{-6.9}          & \textbf{-13.2}         & \textbf{-15.2}         & \textbf{-8.0}          & \textbf{-18.3}         & \textbf{-19.2}         \\ 
\hline
\multirow{5}{*}{D}     & BasketballPass      & -2.4                               & -1.0                               & -3.5                               & -7.1                               & -9.2                               & -13.5                              & -8.0                   & -8.7                   & -15.7                  & -8.8                   & -18.5                  & -18.0                  & -8.7                   & -14.5                  & -18.4                  & -9.8                   & -19.6                  & -22.1                  \\  
                       & BQSquare            & -2.0                               & 0.2                                & -3.7                               & -5.3                               & -2.0                               & -6.0                               & -6.4                   & -0.5                   & -5.3                   & -6.8                   & -4.6                   & -9.5                   & -8.9                   & -3.0                   & -14.5                  & -9.9                   & -6.1                   & -17.1                  \\  
                       & BlowingBubble       & -2.0                               & -0.5                               & -2.8                               & -4.9                               & -5.9                               & -5.2                               & -5.3                   & -4.2                   & -6.6                   & -5.9                   & -11.5                  & -11.2                  & -6.3                   & -9.2                   & -9.8                   & -7.4                   & -14.8                  & -13.6                  \\  
                       & RaceHorses          & -2.5                               & -2.4                               & -3.5                               & -6.1                               & -8.7                               & -12.1                              & -5.9                   & -6.1                   & -11.7                  & -6.4                   & -12.8                  & -15.3                  & -6.6                   & -11.9                  & -13.7                  & -7.6                   & -17.4                  & -17.4                  \\  
                       & \textbf{Average}    & \textbf{-2.2}                      & \textbf{-0.9}                      & \textbf{-3.4}                      & \textbf{-5.8}                      & \textbf{-6.5}                      & \textbf{-9.2}                      & \textbf{-6.4}          & \textbf{-4.9}          & \textbf{-9.8}          & \textbf{-7.0}          & \textbf{-11.9}         & \textbf{-13.5}         & \textbf{-7.6}          & \textbf{-9.6}          & \textbf{-14.1}         & \textbf{-8.7}          & \textbf{-14.5}         & \textbf{-17.5}         \\ 
\hline
\multirow{4}{*}{E}     & FourPeople          & -3.1                               & -1.6                               & -1.6                               & -7.2                               & -5.5                               & -5.6                               & -8.3                   & -4.1                   & -5.9                   & -9.3                   & -9.6                   & -9.5                   & -8.0                   & -10.6                  & -11.3                  & -9.6                   & -14.9                  & -14.5                  \\  
                       & Johnny              & -2.0                               & -1.7                               & -3.1                               & -6.3                               & -9.4                               & -8.7                               & -8.2                   & -7.0                   & -10.5                  & -9.4                   & -15.0                  & -13.1                  & -8.3                   & -15.6                  & -15.0                  & -9.5                   & -20.3                  & -16.7                  \\  
                       & KristenAndSara      & -2.6                               & -1.5                               & -1.7                               & -6.4                               & -8.0                               & -7.3                               & -7.4                   & -4.6                   & -8.6                   & -8.2                   & -11.3                  & -12.0                  & -7.8                   & -14.7                  & -12.6                  & -9.0                   & -19.5                  & -15.3                  \\  
                       & \textbf{Average}    & \textbf{-2.6}                      & \textbf{-1.6}                      & \textbf{-2.1}                      & \textbf{-6.6}                      & \textbf{-7.7}                      & \textbf{-7.2}                      & \textbf{-7.9}          & \textbf{-5.2}          & \textbf{-8.3}          & \textbf{-8.9}          & \textbf{-12.0}         & \textbf{-11.5}         & \textbf{-8.0}          & \textbf{-13.6}         & \textbf{-13.0}         & \textbf{-9.4}          & \textbf{-18.2}         & \textbf{-15.5}         \\ 
\hline
\multicolumn{2}{c}{\textbf{All}}           & \multicolumn{1}{l}{\textbf{-1.6}} & \multicolumn{1}{l}{\textbf{-1.9}} & \multicolumn{1}{l|}{\textbf{-2.9}} & \multicolumn{1}{l}{\textbf{-4.9}} & \multicolumn{1}{l}{\textbf{-6.7}} & \multicolumn{1}{l|}{\textbf{-8.2}} & \textbf{-5.8}          & \textbf{-4.5}          & \textbf{-9.7}          & \textbf{-6.7}          & \textbf{-12.6}         & \textbf{-14.5}         & \textbf{-7.0}          & \textbf{-10.7}         & \textbf{-12.7}         & \textbf{-8.3}          & \textbf{-15.8}         & \textbf{-16.2}         \\ 
\hline
\hline
\end{tabular}

}
\end{table*}




Table \ref{tabbdr2} presents the performance of our proposed method against the anchor compared with the two benchmark methods. It can be seen that in the CTC QP range, the proposed method can achieve an average BD-rate gain of 6.7\%, 12.6\% and 14.5\% on Y, U and V components, respectively. In the same QP range, it is also observed that the proposed method with the prediction signal outperforms the proposed method without the prediction signal by 0.9\%, 8.1\% and 4.8\%, on Y, U and V components, respectively. Compared to the other two JVET solutions, the proposed method shows a significant gain, in the CTC QP range. 

At high QP range, where artifacts are significantly stronger, the only comparison is between the proposed method with and without the prediction signal. As can be seen in Table~\ref{tabbdr2}, the proposed method can achieve an average BD-rate gain of 8.3\%, 15.8\% and 16.2\% on Y, U and V components, respectively. Same as CTC QP range, the use of the prediction signal in high QPs also further increases the gain with an average BD-rate of 1.3\%, 7.1\% and 3.5\% on Y, U and V components, respectively.

In both QP ranges, the achieved BD-rate gain of using the prediction signal is relatively higher for the U and V components than for the Y component. This can be due to the fact that in VVC, there are advanced tools for chroma coding to exploit the redundancies. Examples of such tools are Luma Mapping with Chroma Scaling (LMCS), Joint Cb-Cr residual coding (JCCR), Cross-Component Linear Modeling (CCLM) and a specific chroma IPM called luma Derived Mode (DM) \cite{vvc}. The use of coding information such as intra prediction might enable the CNN-based QE to benefit from the existing correlations and more efficiently predict the compression artifacts. 




\section{Conclusion}
\label{sec:con}
\noindent
In this paper, a CNN-based quality enhancement method was proposed for VVC coded frames, that benefits from the coding information in the intra prediction signal of each frame. The experiments showed that using prediction information can significantly improve the performance of the CNN-based enhancement methods, both for luma and chroma components of intra frames. 
The best explanation for the observed improvements is that exposing the CNN training process to coding information of the sequences, along with their ground-truth original signal, helps them is learning the pattern of compression artifacts. Hence, when the networks are used for the QE task of actual compressed sequences, they can more efficiently recover the lost information.

\bibliographystyle{unsrt}
\bibliography{mybib}

\end{document}